\documentclass[journal]{IEEEtran}

\usepackage{amsmath,amssymb,amsfonts}
\usepackage{amsthm}
\usepackage{cite}
\usepackage[pdftex]{graphicx}
\usepackage[caption=false,font=footnotesize]{subfig}
\usepackage{verbatim}
\usepackage{array}
\usepackage{algorithm}
\usepackage{stfloats}
\usepackage{url}
\usepackage{caption}
\usepackage{color,soul}

\begin{document}

\title{MIMO Operations in Molecular Communications: Theory, Prototypes, and Open Challenges}

\author{Bon-Hong, Koo, Changmin, Lee, Ali E. Pusane,~Tuna Tugcu,~and~Chan-Byoung~Chae\thanks{B.-H. Koo, C. Lee, and C.-B. Chae are with Yonsei University, Korea. A. E. Pusane and T. Tugcu are with Bogazici University, Turkey.}}
\maketitle

\begin{abstract}
The Internet of Bio-nano Things is a significant development for next generation communication technologies. Because conventional wireless communication technologies face challenges in realizing new applications (e.g., in-body area networks for health monitoring) and necessitate the substitution of information carriers, researchers have shifted their interest to molecular communications (MC). Although remarkable progress has been made in this field over the last decade, advances have been far from acceptable for the achievement of its application objectives. A crucial problem of MC is the low data rate and high error rate inherent in particle dynamics specifications, in contrast to wave-based conventional communications. Therefore, it is important to investigate the resources by which MC can obtain additional information paths and provide strategies to exploit these resources. This study aims to examine techniques involving resource aggregation and exploitation to provide prospective directions for future progress in MC. In particular, we focus on state-of-the-art studies on multiple-input multiple-output (MIMO) systems. We discuss the possible advantages of applying MIMO to various MC system models. Furthermore, we survey various studies that aimed to achieve MIMO gains for the respective models, from theoretical background to prototypes. Finally, we conclude this study by summarizing the challenges that need to be addressed.
\end{abstract}

\section{Introduction}

The future of beyond 6G and 7G (B6G/7G) networking is expected to enable novel technologies related to health monitoring, where nanoscale networking has emerged as a promising communication paradigm. Nano-communications  primarily constitutes  the implementation of nanoscale devices and their utilization for \textit{in vivo} applications. On the other hand, recent studies show that electromagnetic waves may not be compatible with nano-communication systems. Techniques that use chemical particles as substitute messengers are therefore of interest for communications technologies.
	A macroscopic view of the trajectory and possible future directions of molecular communications (MC) are presented in~\cite{Farsad2016}. System design with analytical formulations~\cite{Jamali2019} and testbed studies~\cite{Lee2020,Guo2020} are both ongoing with open problems and challenges.
	The main challenge in this field over the next decade is expected to be technology implementation with higher data rates and lower error rates. This article discusses the tasks and challenges of applying multiple-input multiple-output (MIMO) to MC in theory and prototypes, as well as research directions for contemporary researchers.

Particle behaviors are dependent on diffusion and drift; thus, increasing the data rate is a crucial challenge for implementing the MC protocols. In radio frequency (RF) communications, the standard strategy to increase data rate is to obtain independent information paths by exploiting the resources of power, time, frequency, and space. An analogous strategy has also been used in MC; molecular type can be one another resource in MC. It is noteworthy that earlier studies on MC focused more on power, time, and type-based modulation techniques, whereas spatial resources have been studied in recent years~\cite{Farsad2016}.

In~\cite{Kim2013}, the authors introduced modulation techniques to exploit the variable resources of molecular types and power. They claimed that both type- and power-based modulation techniques could either increase the data rate or reduce errors, and that a type domain ensured performance gain while remaining within the same power constraints. The results were numerically validated.
The authors in~\cite{Farsad2018} proposed and compared time-based modulation techniques with power modulation. Owing to the high statistical noise from diffusion, the time domain was shown to yield performance gains in both modulations, and their testbed confirmed the results.

It is essential to investigate MIMO configurations to utilize spatial domain resources. In MC with a diffusion environment, it can be argued that the MIMO operation may not work as in RF communication because of the completely different channel characteristics. However, the authors in~\cite{Koo2016} showed that it is feasible to achieve performance gains in MC by applying MIMO systems in practice. They proposed novel detection algorithms for MC with a 2$\times$2 MIMO configuration and implemented a prototype to test the results. Lee \textit{et al.}~\cite{Lee2017} designed learning-based algorithms to cope with unknown channel state information (CSI), in addition to developments in~\cite{Koo2016}.
Yu \textit{et al.}~\cite{Huang2019} studied an MC system with a 4$\times$4 configuration and demonstrated performance enhancements. The authors in~\cite{Rouzegar2019,Cong2020} investigated general configurations and considered mobility and asymmetry, respectively. To address the interference, both studies successfully applied a zero forcing detector.
Damrath \textit{et al.} provided analytic discussions on the array gain in MC~\cite{Damrath2018}. These studies showed promising results wherein a spatial domain could result in performance gains. Therefore, we are motivated to find open problems in this research area, whereas unknown CSI and interference could pose technical challenges.

The remainder of this article is organized as follows. We first provide an overview of the MIMO-embedded MC system, including communication models, technical hurdles, and testbed designs. Next, existing solutions are discussed. Thereafter, we discuss the potential of molecular MIMO and conclude this article by summarizing the open challenges.

\begin{figure*}[!t]
	\centering
 \includegraphics[width=1.75\columnwidth,keepaspectratio]{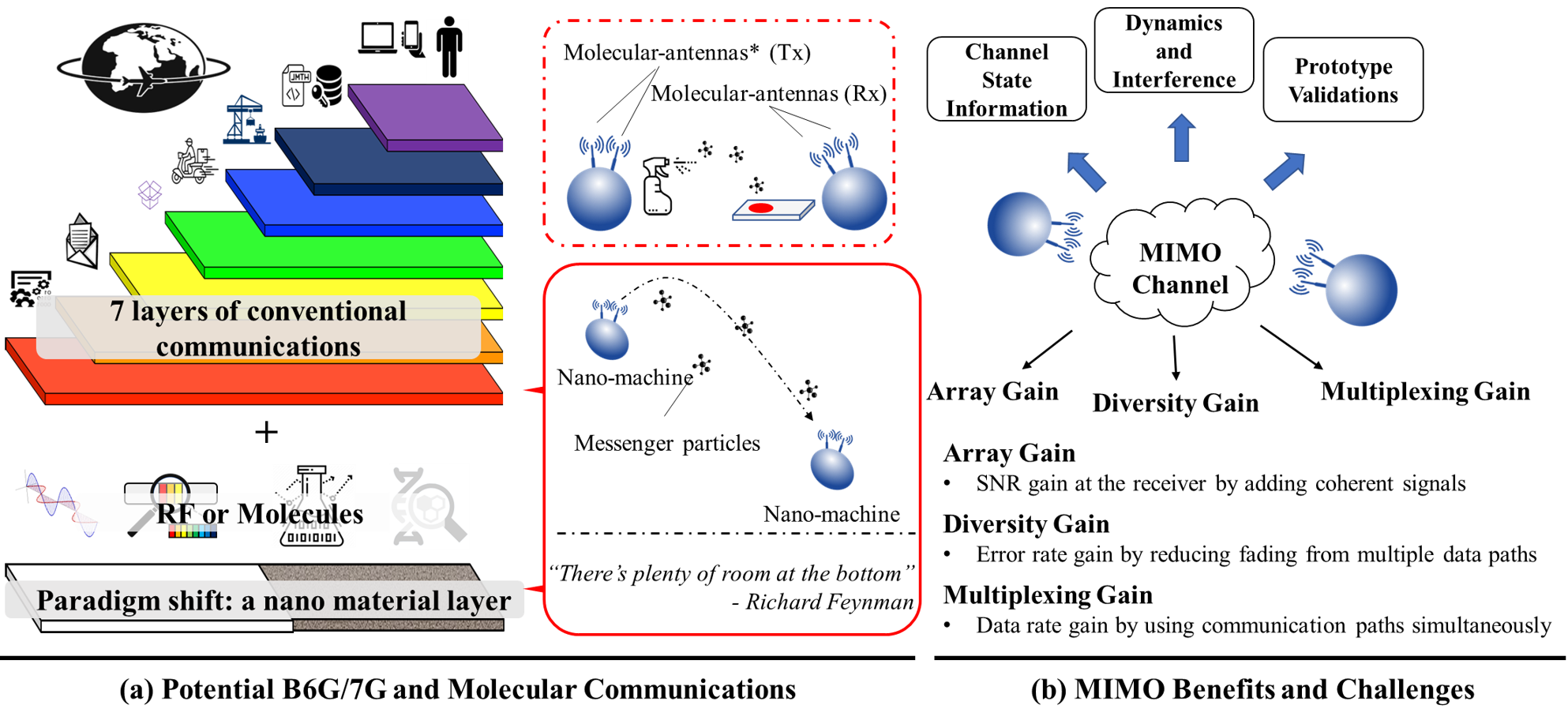}
	\caption{Conceptual overview of B6G/7G and molecular MIMO communications.}
	\label{fig_6G_MIMO_overview}
\end{figure*}

\section{Applying MIMO to MC}
\label{Sec_MC_MIMO}
As shown in Fig. 1, we expect the research area of MC to augment the B6G/7G communication paradigm. In addition to the other MC techniques, we believe that the role of MIMO technologies deserves special consideration. We briefly introduce the potential gains of MIMO for RF communications and review the applications of MC in various scenarios.

\subsection{Brief Review of MIMO Gain in RF Communications}
In RF communications, multiple antennas can provide performance benefits via array, diversity, and multiplexing gains. Array gain is the benefit of having a high signal-to-noise ratio (SNR) at the receiver, which can be achieved by coherently adding received signals at the antennas. A transmitter is required to have CSI for signal coherence. Diversity gain offers a low error rate by providing independent channel paths and reducing the fading effects.
In contrast, multiplexing can achieve an increase in the data rate using simultaneous data streams through pairs of transmitter-receiver antennas. This technique is also referred to as spatial multiplexing.

Knowledge on channel models, signal coherence, and antenna correlations is required to apply MIMO technologies. In MC, a standard channel model is still in the development stage. Furthermore, a diffusion-dominated environment makes it difficult to achieve coherence. Therefore, channel, interference, detection, and verification are prerequisites for applying MIMO schemes to MC. The term 'antenna' for MC in this article can physically be chemical/biological transceivers (molecular emitters and detectors).

\subsection{MC System Model}

We consider three models of interest as depicted in Fig.~2: three-dimensional (3D) free-space, bounded space, and unestablished channel models (e.g., underwater). The first column shows the free-space channel model, where molecules propagate through diffusion and drift. A possible target scenario for the use of this model is tactical nano-robots because of its strength against eavesdropping and energy efficiency. The second column presents a vessel-bounded channel, that mimics \textit{in vivo} applications. Health monitoring by nano-robots with medical sensors and theranostics is a potential application for this model. The third column describes underwater communication, in which RF signal experiences severe degradation. In the bottom row, the corresponding state-of-the-art testbeds are introduced. The first two testbeds are equipped with 2$\times$2 MIMO configurations~\cite{Lee2017,Lee2020} while the third ~\cite{Guo2020} is equipped with a single-input single-output (SISO) configuration.

Mimicking nature is one of the strong motivations for MC technology. For instance, ants in a colony communicate with each other by pheromone molecules, bio-organs in a body organize the system with hormone chemicals, and a shark detects its target from a long distance with a minute amount of odor molecules. From these examples, the following directions can be considered: increasing the number of communication units that a network comprises, expanding the size of a set of chemical compounds that the system employs, and enabling position tracking functions with higher sensitivity.

	 In this article, we base our discussion on a 2$\times$2 symmetric configuration for simplicity. Generalized designs are presented in the following section. The transmitter antennas are regarded as point sources that release messenger particles into the channel medium. The receiver antennas were spherical in shape for analytical tractability. These antennas absorb the contacted molecules and count their numbers over time. In addition to the statistical counting noise, additional noise is considered to account for the unmodeled phenomena. It is assumed that the transceivers are synchronized with one another and are aware of the predefined system parameters, including the required CSI and symbol duration.

\begin{figure*}[!t]
	\centering
	\includegraphics[width=1.75\columnwidth,keepaspectratio]{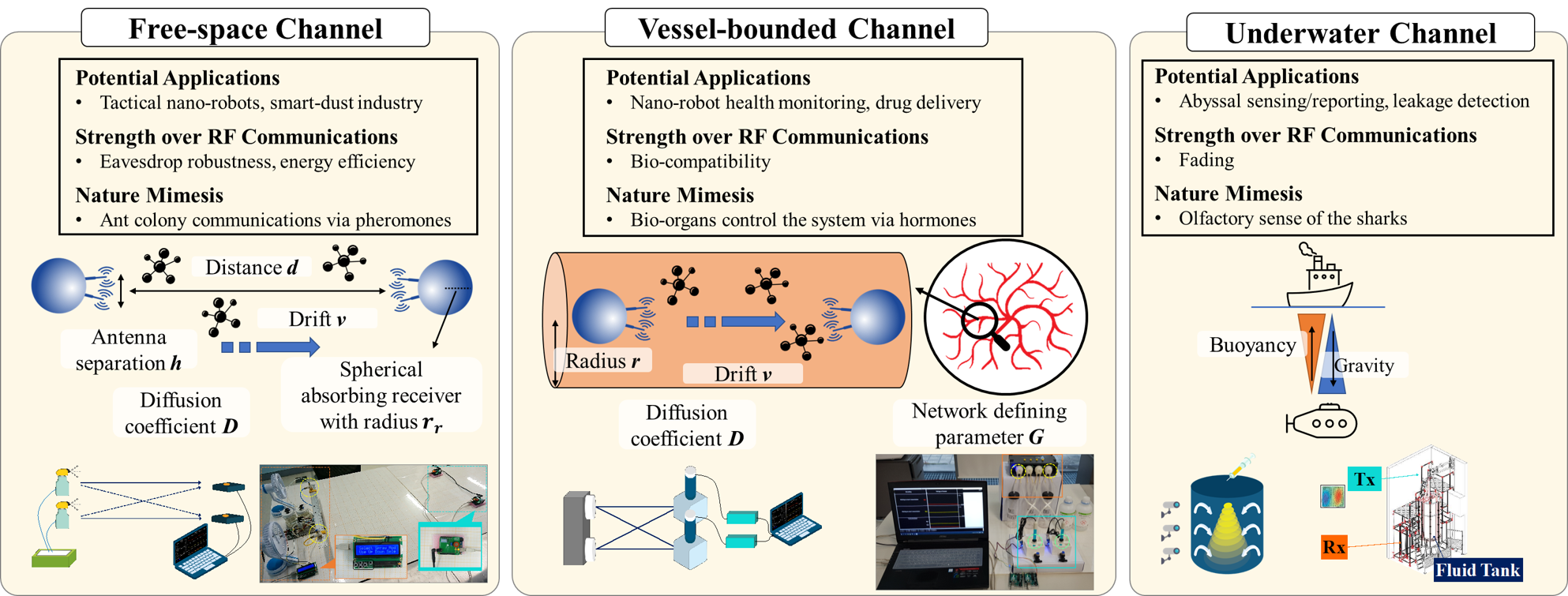}
	\caption{Three types of channel model and comparisons in MC MIMO, from theory to prototypes.}
	\label{fig_ChannelModels}
\end{figure*}

\subsection{Channel Model and CSI}

Free space diffusion dynamics inherently have an indeterminacy of exact movement, and analysis based on statistical characteristics is an alternative approach to address this issue. In this regime, the channel that the information carriers experience is the expected fraction of the number of molecules that the receiver captures with respect to time. A closed-form channel model solution for a case without drift and multiple antennas is presented in~\cite{Yilmaz2014}. Furthermore, Jamali \textit{et al.} presented solutions for a channel with drift and a non-absorbing receiver~\cite{Jamali2019}.

In MIMO configurations, molecular absorption at one receiver antenna induces dependency on the others, and finding a closed-form channel model remains an open problem. The studies presented in~\cite{Koo2016,Lee2017} revealed that it is possible to closely approximate the MIMO channel model by tuning the parameters using a SISO solution.

The CSI required for the transmitter in this model comprises the drift velocity, diffusion coefficient, distance, and MIMO parameters. The challenge is to tune MIMO parameters for each topology, where the relational formula has not yet been solved.
The requirement of retuning parameters for varying topologies makes the system sensitive to displacements of the communication units. Machine learning-based parameter tuning algorithms were introduced in~\cite{Lee2017}, which manage the adjustment of the parameters given the communication distances in the provided training data.

We consider \textit{in vivo} applications, such as the communication units located in blood vessels, where the dominance of the drift effect is high and boundary restrictions affect the channel. The vessel-bounded channel space is regarded as a straight cylinder~\cite{Jamali2019,Koo2016,Farsad2018}. Threshold-based detection algorithms were proposed in~\cite{Koo2016}, and learning-based algorithms were presented in~\cite{Farsad2018} for SISO configurations. In vessel-bounded MC, the set of channel state parameters consists of the distance, diffusion coefficient, velocity profile, and vessel radius.

One research direction for MC channels is the transition from mathematical analysis to prototype implementations. In~\cite{Jamali2019}, the authors considered advective flow and chemical reaction kinetics to provide vessel-bounded MC channel responses. However, this becomes increasingly difficult as the channel becomes more complicated, and another research approach is required.

An alternative approach of studying channels begins from a testbed with numerous measurements and then involves determining {correlations} between sets of input and output, as in~\cite{Lee2020,Koo2020}. The authors in~\cite{Lee2020} discovered that the velocity profile is related to the radius, and both studies showed communication feasibility in vessel-bounded MC. Challenge in these research directions is that prototypes must be developed beforehand.

Underwater environments with severe RF signal degradation are among the prospective fields for the application of MC.
The olfactory sense of sharks outperforms any other alternative communication for the subaqueous positioning task; thus, underwater MC is desirable in addition to acoustic/RF communications~\cite{Guo2020}.
The aforementioned channel models may not be feasible in these environments. It is highly recommended to use a network with multiple devices and antennas for long-distance communication. Therefore, it is desirable to design communication transceivers with multiple components and limited channel knowledge.

\begin{figure*}[!t]
	\centering
	\includegraphics[width=1.8\columnwidth,keepaspectratio]{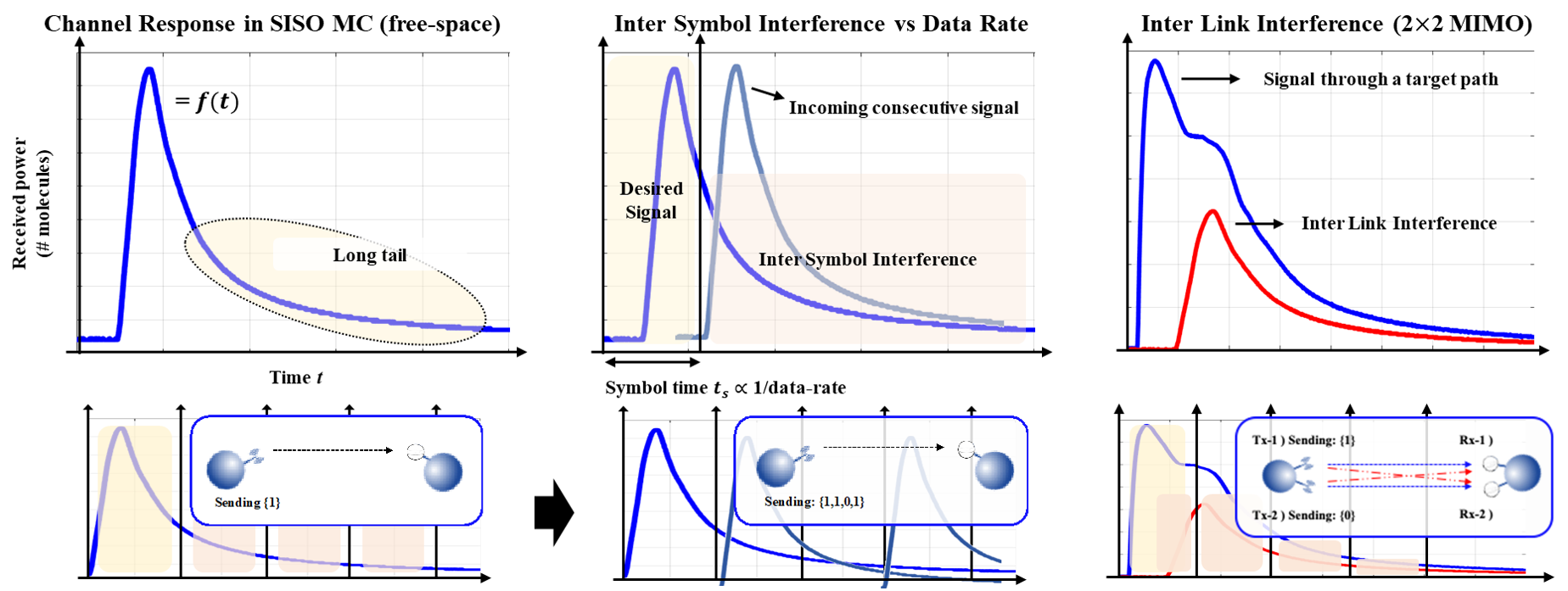}
	\caption{ISI and ILI models for MIMO operations in MC.}
	\label{fig_InterferenceDescriptions}
\end{figure*}

\subsection{Interference Sources and Mitigation Strategies}
Diffusion-based MC suffers from statistical noise as a consequence of Brownian motion. Particles that fail to reach their target within the desired time slot are regarded as interference. Therefore, acquiring a higher data rate comes at the cost of interference. As evident in {the first column of} Fig. 3, the average receiver response to the single-shot transmission has a long tail, which implies that MC suffers from severe interference and noise.

In SISO MC, the molecules that cannot arrive within the symbol duration are the only interference sources, and are analogous to the inter-symbol interference (ISI) in RF communications. The second column of Fig. 3 shows the effect of a long tail on successive signal responses.
Removing remnant molecules with appropriate enzymes can mitigate interference while increasing the system costs. Otherwise, the best available strategy to date is to exploit the statistical channel characteristics and optimize the symbol duration. Kim~\textit{et al.} proposed solutions to achieve a high data rate given the constraints on the SNR level~\cite{Kim2013}.

Applying MIMO schemes to MC yields additional spatial resources for the system. Attaining an array or diversity gain by scheduling the transmitter antennas and summing the receiver antenna signals is straightforward. Koo \textit{et al.} reported that achieving multiplexing gain is possible when inter-link interference (ILI) is considered~\cite{Koo2016}. When the antenna of a device is paired with the closest antenna of the communication counterpart, molecules at the wrong destination are regarded as ILI, as shown in the last column of Fig. 3. Note that the curves in Fig. 3 are averaged values obtained from numerous observations at the receiver, whereas the actual receptions are random variables. In~\cite{Jamali2019}, a Poisson random channel model was introduced to describe the actual responses.

Based on an MC feature where the signal degradation as a function of range is greater than that in RF communications, a simple idea for mitigation is to consider ILI noise and apply a SISO threshold. The knowledge of ILI characteristics results in even better performance, as shown in~\cite{Cong2020} with the equalization method inspired by the zero-forcing method. Koo \textit{et al.} also proposed zero-forcing detection algorithms using statistical characteristics only at the receiver under the assumption of partial CSI available at the receiver~\cite{Koo2016}. The authors in~\cite{Rouzegar2019} also exploited the mean value of the channel and interference and proposed zero-forcing, minimum mean square error, and least square detectors. They also showed that having decision feedback at the transmitter can achieve enhanced performance by eliminating successive interference.

A scenario with multiple transmitters and receivers also introduces an inter-user interference (IUI). IUI can be regarded as an additional ILI, and additional spatial multiplexing gain can be exploited when pairs of communications are predefined.

As the medium drift speed increases, the ratio of ILI to the desired signal decreases. In vessel-bounded channels, where drift dominates over diffusion, the ILI is negligible when the antenna configuration is well aligned in a single stream. Thus, SISO detection algorithms can mitigate ISI for the MIMO setup. Example detection algorithms are {provided} in~\cite{Koo2020}, where it can be observed that the sensor's chemical instability can also yield heavier tails.

We can also assume that communication units spread over multiple vessel streams, which could exacerbate the ILI (or IUI) effect. Lee \textit{et al.} introduced a vessel testbed that used MC MIMO with intersecting vessels and empirically demonstrated that multiplexing gain is achievable with CSI at both the transmitter and receiver~\cite{Lee2020}. The issue of modeling and solving complex vessel topologies is an open problem.

\section{Communication Transceiver Design}

In this section, MC modulation techniques are presented, their capabilities for MIMO applications are discussed, and we also introduce detection algorithms.

\subsection{Modulations in MC}
Modulation schemes can exploit, several domains simultaneously such as time, molecular type, power, and space. The power in MC is proportional to the number of molecules that are relevant to the concentrations. Power-based modulation in MC is known as concentration shift keying (CSK), which resembles to power amplitude-shift modulations in classical communications. The binary modulation of CSK (BCSK) maps bit-1 into a fixed number of molecules and bit-0 into zero molecules. Sending zero molecules is not mandatory; however, it is a popular option in terms of power and accuracy. Therefore, BCSK is used interchangeably with on-off keying (OOK) in MC. A high order of modulation is achievable using multiple molecular quantity levels.

The next common modulation source is the timing of molecular emission. One such example is pulse position modulation (PPM)~\cite{Gursoy2019}. The system has a fixed symbol time, which is divided into multiple time slots. A single-shot at one of the slots per symbol time was used, and the slot position represented a data symbol. Another example in~\cite{Farsad2018} employed single-slot timing modulation, which is equivalent to PPM with an infinite symbol time and indefinite modulation order.

When a transmitter and a receiver can use multiple types of molecules, additional degrees of freedom can be achieved by the type domain. In~\cite{Kim2013}, the authors proposed molecular type-based modulation. Each molecule represents a different symbol. Assuming that the molecules do not interact with each other, the type domain is independent of the others. Therefore, this domain is typically used in combination with other domains. While the domain has unlimited potential, we need to be cautious when planning to use the type domain because the complexity growth is not yet quantifiable and some of the chemicals may not be eligible for use in practice.

A device with multiple antennas yields another independent domain, thereby contributing to the performance gains of the system. In~\cite{Gursoy2019}, the authors proposed index modulation, where each index denoted one transmitter antenna, and the index represented an individual symbol. Yu \textit{et al.}~\cite{Huang2019} also presented space shift keying (SSK) modulation and designed joint modulations of SSK and CSK for a 4$\times$4 MIMO configuration. Gursoy \textit{et al.} addressed two types of molecules using the index modulation in~\cite{Gursoy2019} and combined PPM with space in their subsequent study.

High-order modulation is achievable by adopting more antennas per unit; however, this increases the ILI and decreases the antenna separations. The order of the modulation is also relevant to the channel and device functions. We assumed an ideal case with $T$ transmitter and $R$ receiver antennas. The maximum number of combinations that can be used for the modulation in theory is $(R+1)^T$, where each transmit antenna can either choose one of the $R$ antennas or not. The combinations of transmitter antennas are preferred to be uncorrelated so that the received signals at each antenna are separable by having high directional transmissions to achieve the optimum.

\begin{figure}[t]
	\centering
	\includegraphics[width=1\columnwidth,keepaspectratio]{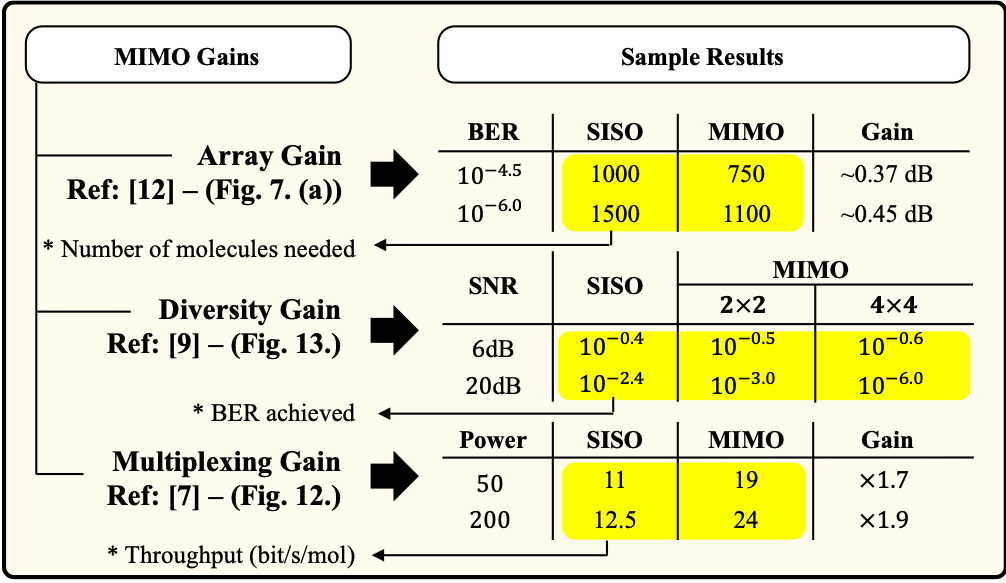}
	\caption{{Applying MIMO to MC increasing performance in all terms: array, diversity, and multiplexing gains.}}
	\label{Fig_DoFGains}
\end{figure}

\subsection{Detection Algorithms}

{As stated in~\cite{Koo2016}, the} best detection algorithm that can be embedded in the receiver is the maximum a posteriori algorithm with perfect CSI. To achieve optimal performance, the knowledge of data statistics, channel governing equations, channel state parameters, and unlimited receiver memory are required. To obtain a practical detection algorithm, the constrained problems must be solved under relaxed conditions. For example, when we have partial CSI, unknown parameters can be estimated, albeit with uncertainty. Assuming that a dataset is non-biased, and we have a reliable channel model, techniques from RF communications such as zero forcing, coding combining~\cite{Damrath2018}, and the Viterbi algorithm properly adapted to the MC environment can be applied.

The authors in \cite{Koo2016} proposed and compared several detection algorithms for cases in which a limited CSI set was provided. The objective was to obtain a threshold for binary detection at the receiver for each bit slot. Increasing the amount of CSI leads to improved performance, although communication is feasible even with a modest amount of CSI. Farsad \textit{et al.}~\cite{Farsad2018} proved that learning-based approaches can detect messages from empirical observations without any prior information on CSI. 

The main difference between MIMO and SISO detection algorithms is the presence of multiple streams and the absence of an exact channel model. For example, in pursuing multiplexing gain, it is assumed that the transceiver antennas are paired and handled the ILI by considering it as additional noise. As mentioned previously, this model is feasible under the condition of antenna pairing in the aligned topology. This problem needs to be addressed to mitigate the ILI when the configuration is not aligned, where the ILI can overwhelm the primary signal.

Machine learning algorithms are compatible with MC with an unknown channel model. This approach requires a lengthy measurement period for both prototypes and numerical simulators. Here, the challenge is that a sufficiently large dataset may not be available. To obtain the maximum amount of possible data, device sharing or online data harvesting studies will be required in the future. Furthermore, effective learning algorithms must be developed.

It was demonstrated in~\cite{Lee2020} that knowledge of the channel can reduce the burden of training, making it possible to train on a smaller training dataset. Farsad \textit{et al.} stated that ensuring adequate memory at the receiver enables recurrent neural networks (RNNs) to perform better. Lee \textit{et al.} presented a learning algorithm for molecular MIMO that trains MIMO channel tuning parameters in combinations or independently. Therefore, further investigation of the channel considering the developments in training neural networks is required.

\subsection{Exploitation of Spatial Resources}
The number of paths that can be created using $T$ transmitter antennas and $R$ receiver antennas is $T \times R$. However, the antenna correlation limits performance gains; the number of uncorrelated signals at the receiver is bounded between $R$ and one. The MC channel attenuates highly over distance; thus, it is sensitive to physical separation. Misalignment can also affect the effectiveness of receivers.

To the best of our knowledge, spatial {correlation} studies between diffusive signals have not yet been conducted. Antenna {correlation} studies can provide an upper bound of the gains for general configurations and help decide the order of priority to explore. Therefore, it is expected that characterizing the correlation effects for transmitter and receiver antennas can result in further progress in the theory and prototyping of molecular MIMO systems.

The effectiveness of the receiver can be further enhanced by ensuring the transmitter antenna directivity. The number of uncorrelated signals at the transmitter antennas depends on their correlation. It is easier to adjust the transmitter antennas than the receiver antennas, because the presence of a single transmitter antenna does not affect the others.

After obtaining the information paths, performance gains can be achieved by applying MIMO strategies as in RF communications. The authors in~\cite{Damrath2018} reported that array gain is achievable by adding coding and receiver combining techniques with the aforementioned detection algorithms, including threshold-finding and learning-based algorithms. The numbers below SISO and MIMO in the top part of Fig. 4 indicate the number of molecules required to achieve the target accuracy.
They proposed that the Alamouti-type code or repetitive-code benefits the system, and a weighted combination of signals at the receiving antennas introduces array gains of 3 dB with a 2$\times$2 MIMO configuration. Their results also indicate that it is feasible to exploit resources for diversity gains in MC MIMO. 

We present the results from~\cite{Huang2019} in the middle part of Fig.~4 to show the diversity gain effects, where the error rates of using SISO, 2$\times$2 MIMO, and 4$\times$4 MIMO systems are compared under the same SNR. The numbers below the configurations represent the symbol error rates.
The results of~\cite{Koo2016} also indicate that multiplexing gain is achievable, as shown at the bottom of Fig. 4. The numbers below the power indicate the number of molecules used to transmit bit-1 while adopting OOK modulation, and the numbers under SISO and MIMO indicate bps per molecule.

\section{Validation by Prototypes}
\label{system_design}

Three types of prototypes {are presented for the three different representative channel types and propagation modes}. The first prototype is based on free-space diffusion and drift. Its state-of-the-art version is equipped with a 2$\times$2 MIMO configuration. We expect its further applications in smart dust and tactical nano-robots. The second prototype is a vessel-bounded testbed. We anticipate that the advanced version will be used in \textit{in vivo} robots where biocompatibility is vital in the future. The third prototype can be operated underwater. This system is capable of supporting futuristic scenarios, including abyssal communications, where RF signal impairments are severe.

\subsection{Free-Space MIMO Testbed}
Communication via free-space propagation is achieved with fans, programmable sprays, and chemical sensors, as shown in the bottom part of Fig. 2. The fan addresses the drift velocity, and its power can adjust the speed. The programmable spray is controlled by an Arduino system on a chip to check the interactive results in a timely manner, while the preprocessed codes are embedded before use. The transmission power is varied by adjusting the size of the spray outlet or spraying duration per shot. The external processors handle the processing of received responses at the sensors; standalone nano-processors can be investigated in future work.

There is wide potential extensibility in the spatial domain, but limited progress has been achieved thus far. Researchers have investigated the variation in performance using 2$\times$2 MIMO and SISO configurations and proved that the results obtained using formulation-based analysis agree well with the prototype results~\cite{Koo2016}. Future upgrade to the free-space MC testbed would focus on increasing the number of adoptable antennas based on suitable theory, while enabling gains from multiple domains.

\subsection{In-Vessel MIMO Testbed}
Researchers have built in-vessel MC systems at multiple scales. Farsad \textit{et al.} designed a meso-scale prototype utilizing base and acid molecules. Programmable pumps introduced molecules into the channel, and pH sensors captured the responses. The desired channel topologies can be expressed using adjustable rubber pipes and joints. The authors applied machine learning for detection, and the RNN significantly enhanced the performance by efficiently mitigating the ISI. It is also worth noting that drift domination results in the sharpness of the peak reactions. This implies high extensibility in the time domain.

Lee \textit{et al.}~\cite{Lee2020} expanded the prototype into a 2$\times$2 MIMO setup. The developed testbed was appropriate for modeling various configurations of tube environments such as the human blood vessel network. A 3D printer was used to print the junction parts.

Koo \textit{et al.}~\cite{Koo2020} implemented a prototype at the nanoscale. Most of the components were analogous to those in~\cite{Lee2020}. One difference was that the sensor was adaptively designed for MC at the nanoscale. In contrast with current nano-sensors, their sensor was rapidly reactive, sensitive to changes in concentration, and inexpensive. Their study is significant for demonstrating the feasibility at the nanoscale and calling for interdisciplinary research for further advancements.

\subsection{Underwater Testbed}

Guo \textit{et al.}~\cite{Guo2020} introduced a macro-scale underwater MC with a prototype that exploited gravity and buoyancy to convey molecules. The objective of this study was to employ MC in liquid environments. In their implementation, two nodes communicated with each other along the vertical axis. The buoy emitted molecules with an initial velocity from the surface, and the molecules were moved by gravity. In contrast, the submarine-unit emitted molecules could float with the help of buoyancy.

Generally, most MC studies in a fluid environment assume laminar flow. However, it is challenging to neglect turbulent flow when the physical size of the application becomes enormous, e.g., tracking the target position beneath the deep ocean. The turbulent flow complicates the analysis of the theoretical channel model, and study on the case is a notable open problem.

\section{Open Challenges and Concluding Remarks}
This study aims to provide insights into the research developments in MC, particularly for exploiting spatial resources with multiple antennas. We surveyed channel models and interference sources in MC and discussed potential technical hurdles that need to be overcome. Next, we summarized the communication strategies considered in MC research and listed open challenges for further study. Because some of the system designs include intractable tradeoffs, it is essential to develop prototypes to test their feasibility and quantify the performances. Therefore, we presented testbed studies for three prospective propagation environments for MC applications. The open challenges identified herein can be summarized as follows:
\begin{itemize}
	\item \textbf{Closed-form channel representations:} Analytic characterizations of channels remain open problems in many cases: free-space propagations with multiple absorbing receiver antennas, vessel-bounded complex environments with junctions and leakages, and undersea propagations.
	\item \textbf{Testbed design for feasibility study:} Various types of hardware should be developed to enable diverse practical applications. Underwater testbeds that perform sensitive position tracking tasks or testbeds with the capability to compose and sense multiple types of chemicals can be one example.
	\item \textbf{Theoretical studies for generalization:} Physical principles that describe the effects of deploying an arbitrary number of antennas are crucial to designing general MIMO communications. For example, knowing the impact of signal correlation between multiple antennas can optimize the MIMO configurations.
\end{itemize}

The current state of MC MIMO research indicates that performance gains from spatial domain exploitation are possible and can be coupled to exploit power, time, and molecular type. In~\cite{Koo2016}, the multiplexing gain results in 1.7 to 1.9 times higher transmission rates with a 2$\times$2 MIMO system when compared with the SISO system. 

Joint optimization techniques with adaptive coding, modulation, and receiver combining schemes can help achieve further efficiency. Joint modulations were proposed and tested virtually in~\cite{Huang2019,Gursoy2019}. We suggest two research directions for further studies. The first involves determining a theoretical bound by comparing the results of the signal correlations for the generalized configurations and modulation orders. This will determine the capacity of the system and provide researchers with insights regarding the areas where research must be prioritized. The other direction is a prototype investigation to obtain quantitative information to deploy multiple antennas in practice.

We foresee the potential of nano-communications in various scenarios. We also anticipate problems in deploying communication systems in new and unexplored environments. The findings of previous studies suggest that learning-based approaches can address unknown channels, provided that we have sufficient observation data. Machine learning is a widely expanding research field, and synergetic effects can be achieved when the learning structure and hardware are specialized for target applications. Hence, interdisciplinary research can boost nano-communication performance, while an optimized learning network is desired beforehand. Hence, we expect that MC hardware, training networks, and transceiver design will be jointly optimized in the future.

\section*{Acknowledgment}
This work was supported in part by the Scientific and Technical Research Council of Turkey (TUBITAK) under Grant 119E190 and part by the Basic Science Research Program through the National Research Foundation of Korea funded by the Ministry of Education (NRF-2020R1A2C4001941).

\bibliographystyle{IEEEtran}
\bibliography{IEEEabrv,MIMO_Opreations_in_Molecular_Communications}

\begin{thebibliography}{10}
\providecommand{\url}[1]{#1}
\csname url@samestyle\endcsname
\providecommand{\newblock}{\relax}
\providecommand{\bibinfo}[2]{#2}
\providecommand{\BIBentrySTDinterwordspacing}{\spaceskip=0pt\relax}
\providecommand{\BIBentryALTinterwordstretchfactor}{4}
\providecommand{\BIBentryALTinterwordspacing}{\spaceskip=\fontdimen2\font plus
\BIBentryALTinterwordstretchfactor\fontdimen3\font minus
  \fontdimen4\font\relax}
\providecommand{\BIBforeignlanguage}[2]{{%
\expandafter\ifx\csname l@#1\endcsname\relax
\typeout{** WARNING: IEEEtran.bst: No hyphenation pattern has been}%
\typeout{** loaded for the language `#1'. Using the pattern for}%
\typeout{** the default language instead.}%
\else
\language=\csname l@#1\endcsname
\fi
#2}}
\providecommand{\BIBdecl}{\relax}
\BIBdecl

\bibitem{Farsad2016}
N.~Farsad \emph{et~al.}, ``{A Comprehensive Survey of Recent Advancements in
  Molecular Communication},'' \emph{{IEEE} Commun. Surveys Tuts.}, vol.~18,
  no.~3, pp. 1887--1919, Feb. 2016.

\bibitem{Jamali2019}
V.~Jamali \emph{et~al.}, ``{Channel Modeling for Diffusive Molecular
  Communication-A Tutorial Review},'' \emph{Proc. {IEEE}}, vol. 107, no.~7, pp.
  1256--1301, Jul. 2019.

\bibitem{Lee2020}
C.~Lee \emph{et~al.}, ``{Demo: In-Vessel Molecular MIMO Communications},'' in
  \emph{Proc. IEEE WCNC Workshops}, Apr. 2020, pp. 6--7.

\bibitem{Guo2020}
W.~Guo \emph{et~al.}, ``{Vertical Underwater Molecular Communications via
  Buoyancy: Gaussian Velocity Distribution of Signal},'' in \emph{Proc. IEEE
  ICC}, Oct. 2020.

\bibitem{Kim2013}
N.-R. Kim \emph{et~al.}, ``{Novel Modulation Techniques using Isomers as
  Messenger Molecules for Nano Communication Networks via Diffusion},''
  \emph{{IEEE} J. Sel. Areas Commun.}, vol.~31, no.~12, pp. 847--856, Dec.
  2013.

\bibitem{Farsad2018}
N.~Farsad \emph{et~al.}, ``{Capacity Limits of Diffusion-based Molecular Timing
  Channels with Finite Particle Lifetime},'' \emph{{IEEE} Trans. Mol. Bio.
  Multi-Scale Commun.}, vol.~4, no.~2, pp. 88--106, Jun. 2018.

\bibitem{Koo2016}
B.-H. Koo \emph{et~al.}, ``{Molecular MIMO: From Theory to Prototype},''
  \emph{{IEEE} J. Sel. Areas Commun.}, vol.~34, no.~3, pp. 600--614, Mar. 2016.

\bibitem{Lee2017}
C.~Lee \emph{et~al.}, ``{Machine Learning based Channel Modeling for Molecular
  MIMO Communications},'' in \emph{Proc. IEEE SPAWC}, Jul. 2017, pp. 1--5.

\bibitem{Huang2019}
Y.~Huang \emph{et~al.}, ``{Spatial Modulation for Molecular Communication},''
  \emph{{IEEE} Trans. NanoBiosci.}, vol.~18, no.~3, pp. 381--395, Jul. 2019.

\bibitem{Rouzegar2019}
S.~M.~R. Rouzegar \emph{et~al.}, ``{Diffusive MIMO Molecular Communications:
  Channel Estimation, Equalization, and Detection},'' \emph{{IEEE} Trans.
  Commun.}, vol.~67, no.~7, pp. 4872--4884, Jul. 2019.

\bibitem{Cong2020}
C.~Wu \emph{et~al.}, ``{Signal Detection for Molecular MIMO Communications With
  Asymmetrical Topology},'' \emph{{IEEE} Trans. Mol. Bio. Multi-Scale Commun.},
  vol.~6, no.~1, pp. 60--70, Jul. 2020.

\bibitem{Damrath2018}
M.~Damrath \emph{et~al.}, ``{Array Gain Analysis in Molecular MIMO
  Communications},'' \emph{{IEEE} Access}, vol.~6, pp. 61\,091--61\,102, Sep.
  2018.

\bibitem{Yilmaz2014}
H.~B. Yilmaz \emph{et~al.}, ``{Three-Dimensional Channel Characteristics for
  Molecular Communications With an Absorbing Receiver},'' \emph{{IEEE} Commun.
  Lett.}, vol.~18, no.~6, pp. 929--932, Jun. 2014.

\bibitem{Koo2020}
B.-H. Koo \emph{et~al.}, ``{Deep Learning-based Human Implantable Nano
  Molecular Communications},'' in \emph{Proc. IEEE ICC}, Jun. 2020.

\bibitem{Gursoy2019}
M.~C. Gursoy \emph{et~al.}, ``{Index Modulation for Molecular Communication via
  Diffusion Systems},'' \emph{{IEEE} Trans. Commun.}, vol.~67, no.~5, pp.
  3337--3350, May 2019.

\end{thebibliography}
\vskip -1.5\baselineskip plus -1fil
\begin{IEEEbiographynophoto}
	{Bon-Hong Koo} is currently a Ph.D. student at the School of Integrated Technology, Yonsei University, Korea. His research interests include molecular communications, and machine learning.
\end{IEEEbiographynophoto}

\vskip -2.5\baselineskip plus -1fil

\begin{IEEEbiographynophoto}
	{Changmin Lee} is currently a Ph.D. student at the School of Integrated Technology, Yonsei University, Korea. His research interests include testbed implementations.
\end{IEEEbiographynophoto}

\vskip -2.5\baselineskip plus -1fil

\begin{IEEEbiographynophoto}
	{Ali Emre Pusane} is a Professor in the Department of Electrical and Electronics Engineering, Bogazici University, Turkey. His research interests include wireless communications, molecular communications, information theory, and coding theory.
\end{IEEEbiographynophoto}

\vskip -2.5\baselineskip plus -1fil

\begin{IEEEbiographynophoto}
	{Tuna Tugcu} is a Professor in the Department of Computer Engineering, Bogazici University, Turkey. His research interest includes nanonetworking, molecular communications, wireless networks, and IoT.
\end{IEEEbiographynophoto}

\vskip -2.5\baselineskip plus -1fil

\begin{IEEEbiographynophoto}
	{Chan-Byoung~Chae} is an Underwood Distinguished Professor at Yonsei University, Korea. His research interest includes emerging technologies for 5G/6G and molecular communications. He is now an EiC of the IEEE Trans. Molecular, Biological, and Multi-scale Communications and an IEEE Fellow.
\end{IEEEbiographynophoto}

\end{document}